\documentclass[showpacs,prb,twocolumn,dvips,pdflatex]{revtex4}
\usepackage{graphicx}
\usepackage{graphics}
\usepackage{dcolumn}
\usepackage{bm}
\usepackage{amsmath}
\usepackage{amssymb}

\def\al{\alpha }
\def\be{\beta }
\def\ga{\gamma }
\def\de{\delta }
\def\ep{\epsilon }

\def\si{\sigma}

\def\t{\tau }
\def\R{{ R}}
\def\S{{\cal S}}
\def\beqnar{\begin{eqnarray}}
\def\eeqnar{\end{eqnarray}}
\def\beq{\begin{equation}}
\def\eeq{\end{equation}}
\def\bvt{\begin{verbatim}}
\def\evt{\end{verbatim}}
\def\bfig{\begin{figure}}
\def\efig{\end{figure}}
\def\bs{\begin{split}}
\def\es{\end{split}}

\begin{document}
\title{A new methodology for the extraction of anharmonic force constants 
from first principles density functional calculations}

\author{K. Esfarjani }
\affiliation{Department of Physics, Sharif University of Technology,
Tehran, Iran, \\
and Department of Physics, UC Santa Cruz, CA, USA\\}
\author{Harold T. Stokes}
\affiliation{Department of Physics, Brigham Young University,
Salt Lake City, UT, USA}
\date{\today}

\begin{abstract}
A new method for extracting force constants (FC) from first principles is 
introduced. It requires small supercells but very accurate forces.
In principle, provided that forces are accurate enough, it can
extract harmonic as well as anharmonic FCs up to any neighbor shell.
Symmetries of the FCs as well as those of the lattice are used to reduce
the number of parameters to be calculated.
Results are illustrated for the case of Lennard-Jones potential 
where forces are exact and FCs can be calculated analytically, 
and Si in the diamond structure.
The latter are compared to previously calculated 
harmonic FCs.
\end{abstract}

\pacs{aaa}
\maketitle

\title{First-principles-based calculation of anharmonic force constants}

\section{Introduction}

Two methodologies have been so far developed to calculate force constants (FC)
of bulk crystals. One relies on perturbation theory and linear response
formalism\cite{baroni}, in which the infinitesimally small 
displacements of atoms are taken as the perturbing potential, and the
required properties, namely the FCs, are calculated as a function of ground 
state eigenvalues and eigenstates of the crystal by using perturbation theory.
The other more traditional method, called the direct 
method\cite{kunc,parlinski,marcel}
requires a small cartesian displacement of an atom placed in a supercell 
(preferably twice larger than the cutoff range of the FCs). 
Forces applied to other atoms divided
by the small displacement produce the required FCs. This method, however
needs also the calculation of Born charges in the case of polar semiconductors,
so that FCs due to long-range Coulombic forces can be added analytically 
to the obtained values.
Anharmonic FCs of third or higher order can also be calculated in the
same fashion. In the direct method, in principle two small displacements along 
a given direction are needed in order to fit the produced force with 
a second order polynomial and extract the harmonic and cubic coefficients. 
However, we are not aware of any calculation involving three-body
(or higher) interactions.
Perturbation theory, on the other hand, uses the so-called ``2n+1 formula" 
\cite{gonze}
to get FCs up to order 2n+1 from a $n^{\rm th}$ order perturbation expansion
of the electronic wavefunctions.
One can see that the direct method, not being a very systematic one, soon
becomes prohibitive if FCs of third or fourth order are needed, given their
very large number. The calculation of fourth order, or so-called ``quartic" 
terms becomes also quite involved using the perturbation theory method 
as now wavefunctions need to be expanded up to second order.

In this paper, we propose a  methodology to extract the harmonic as well 
as cubic and quartic anharmonic FCs from first-principles calculations
in a systematic way.
While harmonic FCs are used for the calculation of phonon spectra, 
vibrational modes and elastic properties, cubic anharmonic FCs give 
the phonon lifetimes
and scattering rates, and quartic ones correct the shift in the phonon
frequencies. These quantities are the main ingredients in phonon 
transport theories which calculate thermal conductivity.
They can also be used for constructing a classical molecular dynamics 
potential of ab-initio accuracy. 
From such molecular dynamics simulations, thermal conductivity, which 
can be written as the heat current autocorrelation function\cite{kubo} 
and other thermodynamic properties  of bulk or nanostructured materials 
can be calculated.  Phase transformations can also be 
investigated by using anharmonic potentials.

\section{Methodology}

The force constants are defined as derivatives of the potential 
energy with respect to atomic displacements about their equilibrium 
position. We write the potential energy in the following form:
\beqnar
V&=&V_0 + \sum_i \Pi_i u_i +  \frac{1}{2!} \sum_{ij}\Phi_{ij} \, u_i u_j +  \frac{1}{3!} \sum_{ijk}\Psi_{ijk} \, 
u_i u_j u_k \nonumber \\
&+&  \frac{1}{4!} \sum_{ijkl}\chi_{ijkl} \, u_i u_j u_k u_l
\label{potential}
\eeqnar
where the roman index $i$ labels the triplet $(\R,\tau,\al)$ with $\R$ 
being a translation vector of the primitive lattice, $\tau$  an atom within the 
primitive unit cell, and $\alpha$ the cartesian coordinate of the atomic 
displacement $u$. In other words,  $u_i$ is the displacement of the 
atom $(\R,\tau)$ in the direction $\alpha$ from its equilibrium or
any reference position.
$ \Phi,\Psi$ and $\chi$ are respectively the harmonic, cubic and 
quartic FCs, whereas $\Pi$ is the negative of the residual force on atom $i$, 
and is zero if the potential energy $V$ is expanded around its minimum
or equilibrium configuration. 
In clusters or molecules the formalism is the same, only the
translation vector $\R$ needs to be dropped.

The resulting force on atom $i$ would then be:
\beqnar
F_i &=& - \frac{\partial V}{\partial u_i} = -\Pi_i -\sum_j  \Phi_{ij} \,u_j - 
 \frac{1}{2} \sum_{jk}\Psi_{ijk} \, 
 u_j u_k  \nonumber \\ &-&   \frac{1}{3!} \sum_{jkl} \chi_{ijkl} \,  u_j u_k u_l
\label{force}
\eeqnar

If one considers a simple FCC Bravais lattice in 3 dimensions,
one can discover by inspection that there are 123 harmonic, 4867 cubic 
and 95663 quartic anharmonic FCs only within the first neighbor shell.
Of course they are related by different symmetry properties, which
will be used here to reduce their number to respectively to 4,12 and 56.

\subsection{General symmetries of the problem}
The symmetries of the FCs are deduced from 
rotational and translational invariance of the system, in addition to the 
symmetries of the crystal itself. 

These symmetries are:\\ 
{-Invariance under the permutation of the indices:}\\ 
\beqnar
 \Phi_{ij}&=& \Phi_{ji} \nonumber \\
\Psi_{ijk}&=& \Psi_{ikj} = \Psi_{jik} = \Psi_{kji} = ... \nonumber \\
\chi_{ijkl}&=&\chi_{ikjl}=\chi_{ijlk}=\chi_{jikl} = ...
\label{permut}
\eeqnar
where $i,j,k$ and $l$ 
refer to neighboring atoms.
This comes about because the force constants are  
derivatives of the potential energy, and one can change the order of 
differentiation for such analytic function.\\

{-Invariance under an arbitrary translation of the system:}\\
Translational invariance of the whole system (even if not an ordered crystal) 
also implies $V(u) = V(u+c)$ and $F(u)= F(u+c)$ (which is easier to use)
where $u(t)$ are the dynamical variables, and $c$ is a constant arbitrary 
shift vector (may be incommensurate with $\R$).

This implies the well-known ``acoustic sum rule" (ASR) generalized to higher order FCs:
\beqnar
0 &=& \sum_{\t} \, \Pi_{0 \t}^{ \al} \,\,\, \forall (\al) \,\, => {\rm Total \,~force \,~on\,~unit\,~cell = 0}  \nonumber \\
0 &=&  \sum_{\R_1,\t_1} \, \Phi_{0\t,\R_1 \t_1}^{\al\be} \,\,\, \forall (\al\be,\t) \nonumber \\
0 &=& \sum_{\R_2,\t_2} \, \Psi_{0\t,\R_1\t_1,\R_2\t_2}^{\al\be\ga} \,\,\, \forall(\al\be\ga,\R_1,\t_1)\nonumber \\
0 &=& \sum_{\R_3,\t_3} \, \chi_{0\t,\R_1\t_1,\R_2\t_2,\R_3\t_3}^{\al\be\ga\de} \,\,\,  \forall(\al\be\ga\de,\R_1\R_2,\t_1\t_2)
\label{trans}\eeqnar
so that diagonal terms in these tensors can be defined in terms of their 
off-diagonal elements, for arbitrary cartesian components. For more details
about the proof of these relations and the ones following on rotational
invariances, we refer the reader to Ref. \cite{llreview}. 


{-Invariance under an arbitrary rotation of the system:}\\
Likewise if the system is rotated arbitrarily, the total
energy and forces should not change. This leads to the following
relations\cite{llreview}:

\beqnar
0 &=& \sum_{\t} \,  \Pi_{0 \t}^{ \al}\, \t^{\be}\, \ep^{\al\be\nu}, \,\,\, \forall ( \nu) \,\, ({\rm Torque \,on\,unit\,cell}=0) \nonumber \\
0 &=&  \sum_{\R_1,\t_1} \, \Phi_{0\t,\R_1 \t_1}^{\al\be}\, (\R_1\t_1)^{\ga}\, \ep^{\be\ga\nu}
 + \Pi_{0 \t}^{ \be }\, \ep^{\be \al \nu}  \,\,\, \forall (\al\nu,\t)\nonumber \\
0 &=& \sum_{\R_2,\t_2} \, \Psi_{0\t,\R_1\t_1,\R_2\t_2}^{\al\be\ga} \, (\R_2\t_2)^{\de} \, \ep^{\ga\de\nu} +
\Phi_{0\t,\R_1\t_1}^{\ga\be}\, \ep^{\ga\al\nu} \nonumber \\
&+& \Phi_{0\t,\R_1\t_1}{\al\ga}\, \ep^{\ga\be\nu} \,\,\, \forall(\al\nu,\R_1,\t\t_1)
\nonumber \\
0 &=& \sum_{\R_3\t_3}\, \chi_{0\t,\R_1\t_1,\R_2\t_2,\R_3\t_3}^{\al\be\ga\de}\,
(\R_3\t_3)^{\mu}\, \ep^{\de\mu\nu} \nonumber \\ 
& +& \, \Psi_{0\t,\R_1\t_1,\R_2\t_2}^{\de\be\ga} \,\ep^{\de\al\nu} +
 \, \Psi_{0\t,\R_1\t_1,\R_2\t_2}^{\al\de\ga} \,\ep^{\de\be\nu}  \nonumber \\
& +& \, \Psi_{0\t,\R_1\t_1,\R_2\t_2}^{\al\be\de} \,\ep^{\de\ga\nu} 
\,\, \, \forall(\al\nu,\R_1\R_2,\t\t_1\t_2)
\label{rot}
\eeqnar

Here $ \ep^{\al\be\ga}$ is the anti-symmetric Levy-Civita symbol, and
$(\R\t)^{\al}$ refers to the cartesian component $\al$ of the reference position
vector of the atom $\t$ in unit cell defined by $\R$. Moreover, an implicit
summation over repeated cartesian indices is implied. 

As we see, rotational invariance relates the second to the 
third order terms, and the third to the fourth order terms, implying that
if the expansion of the potential energy is truncated after the fourth order 
terms, we need to start with the fourth order terms, and application of 
rotational invariance rules will give us constraints on third and 
second order FCs respectively.  

\subsection{Point / space group symmetries}

The other constraints come from symmetry operations, such as lattice
translation, rotation, mirror
or any symmetry operation of the space/point group of the crystal/molecule 
which leaves the latter invariant. 

Invariance under a translation of the system by any translation 
lattice vector $\R$ implies the following relations:\\
\beqnar
\Pi_{\R\t}^{\al} &=& \Pi_{0 \t}^{ \al} \,\, \forall (\R \t \al) \nonumber \\
\Phi_{\R\t,\R_1\t_1}^{\al\be} &=& \Phi_{0 \t ,\R_1-\R \t_1}^{\al\be} \nonumber \\
\Psi_{\R\t,\R_1\t_1,\R_2\t_2}^{\al\be\ga} &=& \Psi_{0 \t ,\R_1-\R \t_1,\R_2-\R \t_2}^{\al\be\ga} \nonumber \\
\chi_{\R\t,\R_1\t_1,\R_2\t_2,\R_3\t_3}^{\al\be\ga\de} &=& \chi_{0 \t ,\R_1-\R \t_1,\R_2-\R \t_2, \R_3-\R \t_3}^{\al\be\ga\de} 
\eeqnar
So in an ideal crystal, this reduces the number of force constants considerably 
(by the number of unit cells, to be exact), meaning that we will use for the atoms in all the other cells the same FCs
as those we have kept for the atoms in the ``central" cell.

If a rotation or mirror symmetry operation is denoted by $S$, 
we must have:
\beqnar 
\Pi_{S\tau}^{\al} &=& \sum_{\al'} \,\Pi_{\tau}^{\al'} \, \S_{\al,\al'}\nonumber \\
\Phi_{ S\tau, S\tau_1}^{\al\be} &=& \sum_{\al'\be'} \,\Phi_{\tau,\tau_1}^{\al'
\be'} \, \S_{\al,\al'}\, \S_{\be,\be' } \nonumber \\
\label{symm}
\Psi_{ S\tau, S\tau_1, S\t_2}^{\al\be\ga} &=& \sum_{\al'\be'\ga'} \,\Psi_{\tau,\tau_1\t_2}^{\al' \be'\ga'} \, \S_{\al,\al'}\, \S_{\be,\be' } \, \S_{\ga,\ga' }   \\
\chi_{ S\tau, S\tau_1, S\tau_2, S\tau_3}^{\al\be\ga\de} &=& \sum_{\al'\be'\ga'\de'} \,\chi_{\tau,\tau_1,\t_2,\t_3}^{\al' \be' \ga' \de'} \, \S_{\al,\al'}\, \S_{\be,\be' } \, \S_{\ga,\ga'}\, \S_{\de,\de' }  \nonumber
\eeqnar 
where $\S_{\al,\al'}$ are the 3$\times$3 matrix elements of the 
symmetry operation $ S$. 
These symmetry relations impose a {\bf linear} set of constraints on the force
constants. 

We should note that any physically correct model of force constants 
must satisfy the invariance relations. On the other hand, we do 
approximations by truncating the range of FCs and their order in the 
Taylor expansion of the potential. Therefore imposing the constraints
will move their value away from the true value, but has the advantage
that they are physically correct, and will for instance reproduce the linear
dispersion of acoustic phonons near $k=0$. So one should keep in mind
that an unrealistic truncation to a too short of a range will produce 
results in disagreement with experiments.

\section{Implementation}

Given a crystal with its unit cell and atoms, we first identify
its symmetry properties and construct the matrices $\S$. Using the latter,
and the equations \ref{permut} and \ref{symm}, independent FCs of each 
rank are identified. Then a set of force-displacement data calculated from
first-principles is constructed. Next, 
since the data set is (better be) larger than the number of unknown FCs,
the linear set of equations 
\ref{force}, \ref{trans} and \ref{rot} are fitted with this data using
a singular value decomposition (SVD) algorithm. In SVD the  
unknown force constants will be calculated in a least square sense,
and furthermore linear dependencies among the equations will not be a 
problem as the result is projected out of the ``zero eigenvalue space"\cite{nr}.
One has also the option to use the relations \ref{trans} and \ref{rot} to 
eliminate some of the FCs in \ref{force} and solve for the remaining FCs.
Here one needs to make a judicious choice of FCs to be eliminated.
We prefer not to eliminate the FCs and keep the option of checking the 
violation of the translational and rotational invariances if only
relations \ref{force} are used.

The data set can be obtained in several ways: a molecular dynamics run with small
initial displacements, randomly moving all atoms by small displacements a few
time steps and calculating the forces on all atoms, and finally displacing one atom at
a time symmetrically about its equilibrium position by a small displacement along
x,-x,y,-y,z and -z directions and calculating the forces on all other atoms.
Experience has shown that the latter works better for the computation of 
harmonic force constants and two-body force constants in general. For three and
four-body FCs one needs to displace at least  two and 3 atoms at a time, respectively.

We must add that the data set is obtained not from a unit cell but from 
displacements performed in a supercell whose size better exceed the range of FCs;
otherwise the contribution of images from neighboring supercells might also be 
included in a considered FC. In some cases, this could lead to errors in 
the evaluation of FCs.
Notice that the exact outcome of this procedure includes, 
strictly speaking, the contribution of images as well. For instance, 
in the case of harmonic FCs, instead of $ \Phi_{\tau,\tau'}$ actually the 
sum $ \tilde{\Phi}_{\tau,\tau'} = \sum_{L}  \Phi_{\tau,L+\tau'} $, where $L$ 
is a translation vector of the supercell, will be extracted.
Therefore supercells of low symmetry are preferable in order to avoid 
encountering  FCs that can not be determined. For example in a cubic
supercell the force constants between the corners of the cube can never
be calculated this way since the distance between adjacent corners
never changes. It must, furthermore, be emphasized that the size 
of the data set as well as its accuracy are crucial in determining the
correct set of force constants.

\section{Results}

To check the feasibility and accuracy of the method, we first used
the Lennard-Jones (LJ) pair force, for which derivatives can be analytically 
calculated and compared to our results. Furthermore, the LJ forces are 
accurate within the printed number of digits by the computer, and do not suffer
from any convergence or roundoff error problems. First we considered an
FCC-Bravais crystal of LJ particles with integer cartesian coordinates.
Particles were confined to interact only with their first neighbor.
It was found that choosing a very small set of displacements (of the order
of 0.001) produced the best results. Since in both the MD data and the fitting
procedure the cutoff was set so that only first neighbors interact, the SVD procedure
reproduced the exact FCs remarkably well: the error in  the harmonic, cubic and 
quartic FCs was in the 7$^{\rm th}$, 5$^{\rm th}$ and 4$^{\rm th}$ digits respectively. 

As a more stringent test, and in order to get confidence on the accuracy of
the fitting method for a real material such as Si, we also considered a 
diamond structure of LJ particles with interactions up to 10th nearest neighbor, but
limited them in our fitting up to the 8th neighbor only.
The energy unit was taken to be one and the length unit $\si$ was 
taken to be equal to the first neighbor distance ($\sqrt{3}/4$). 
No MD was performed, instead we only displaced the first atom by 0.0005
and 0.001 along the x direction and recorded the forces on all other atoms
in a $3 \times 3 \times 3$ cubic supercell of 216 atoms.
The results are summarized in table \ref{ljfc}.
Three-body forces are found to be respectively 5 and 3 orders of magnitude 
smaller than the largest two-body values in cubic and quartic FCs.  
It is worth noticing that the fitted values obtained  imposing the 
invariance relations are slightly worse that when invariances are not imposed.
This is because some of the longer range FCs (9$^{\rm th}$ and 10$^{\rm th}$ neighbors) 
were neglected in the fitting procedure, and as a result the values of the 
included FCs are slightly affected when invariance relations are imposed. 
In a real material case however, it is more 
important to have physically correct FCs, even though their values might not be exact.
In any case, this can be checked by comparing the results with and without 
imposing invariance constraints. 
Imposing the constraints conditions in the SVD procedure resutls in
the violation of the latter by typically $10^{-8}$, whereas the
force-displacement data could be violated giving relative errors
as large as a few percent, if the chosen cutoff is too short or
first-principles forces are not very accurate, or displacements
are too large.

\begin{table}
\begin{tabular}{c|ccc}
FCs & {Nb-Shell} &  {Exact} & {SVD} \\
\hline
$\Phi^{xx}_{1,1}$ & 0 (0,0,0) & 711.01623  &  711.016    \\
\hline
$\Phi^{xx}_{1,2}$ & 1 (${1 \over 4},{1 \over 4},{1 \over 4}$) & -181.33329  &    -181.333   \\
$\Phi^{xy}_{1,2}$ & 1 & -213.33328  &  -213.333 \\
\hline
$\Phi^{xx}_{1,1}$ & 2 (${1 \over 2},{1 \over 2}$,0) &  1.497986 &  1.49799   \\
$\Phi^{yy}_{1,1}$ & 2 &  -0.5660705 &  -0.56607  \\
$\Phi^{xy}_{1,1}$ & 2 &  2.064056 &    2.06406  \\
$\Phi^{xz}_{1,1}$ & 2 &  0 &   $ 2 \times 10^{-15}$  \\
\hline
$\Phi^{xx}_{1,2}$ & 3 ($-{1 \over 4},-{1 \over 4},-{3 \over 4}$) & -0.0502417  &   -0.050242  \\
$\Phi^{zz}_{1,2}$ & 3 &  0.9066605 &  0.906668  \\
$\Phi^{xy}_{1,2}$ & 3 &  0.1196113 &   0.119612 \\
$\Phi^{xz}_{1,2}$ & 3 &  0.3588383 &   0.358840  \\
\hline
$\Phi^{xx}_{1,1}$ & 4 (1,0,0) & 0.2700770  &  0.270080   \\
$\Phi^{zz}_{1,1}$ & 4 &        -0.0390294  & -0.039029 \\
\hline
$\Phi^{xx}_{1,2}$ & 5 ($-{3 \over 4},-{3 \over 4},-{1 \over 4}$) & 0.05459899  &   0.0545990   \\
$\Phi^{zz}_{1,2}$ & 5 & -0.0114737  &  -0.0114737  \\
$\Phi^{xy}_{1,2}$ & 5 & 0.07433175  &   0.0743320 \\
$\Phi^{xz}_{1,2}$ & 5 & 0.02477725  &   0.0247773 \\
\hline
$\Phi^{xx}_{1,1}$ & 6 (${1 \over 2},{1 \over 2}$,1) & 0.0025635  &  0.0025635  \\
$\Phi^{zz}_{1,1}$ & 6 & 0.0335998  &   0.0336000  \\
$\Phi^{xy}_{1,1}$ & 6 & 0.0103455  &   0.0103454  \\
$\Phi^{xz}_{1,1}$ & 6 & 0.0206909  &   0.0206909 \\
$\Phi^{yz}_{1,1}$ & 6 & 0.0206909  &   0.0206909 \\
\hline
$\Phi^{xx}_{1,2}$ & 7 (${1 \over 4},{1 \over 4},{5 \over 4}$) & -0.00342573  & -0.0034258  \\
$\Phi^{zz}_{1,2}$ & 7 & 0.03109080  &    0.0310910 \\
$\Phi^{xy}_{1,2}$ & 7 &  0.00143819  &   0.0014382 \\
$\Phi^{xz}_{1,2}$ & 7 &  0.00719095  &   0.0071910  \\
\hline
$\Phi^{xx}_{1,2}$ & 7 ($-{3 \over 4},-{3 \over 4},-{3 \over 4}$) & 0.00807978  &   0.0080798 \\
$\Phi^{xy}_{1,2}$ & 7 & 0.01294370  &   0.0129437 \\
\hline
$\Phi^{xx}_{1,1}$ & 8 (1,1,0) & 0.00739133  &   0.0073913 \\
$\Phi^{zz}_{1,1}$ & 8 & -0.00246785  &  -0.0024678  \\
$\Phi^{xy}_{1,1}$ & 8 & 0.00985918  &   0.0098592 \\
$\Phi^{xz}_{1,1}$ & 8 &  0 &   $ 2 \times 10^{-13}$  \\
\hline
\hline
$\Psi^{xxx}_{01,01,02} $ &  & 2673.7772  & 2673.793   \\
$\Psi^{xxy}_{01,01,02} $ &  & 4380.4434  & 4380.590   \\
$\Psi^{xyz}_{01,01,02} $ &  & 5233.7765  & 5234.127   \\
\hline
$\Psi^{xyx}_{01,01,11} $ &  &  4.128112  &   4.1267   \\
$\Psi^{yyy}_{01,01,11} $ &  & -5.453064  &   -5.4645   \\
$\Psi^{yzy}_{01,01,11} $ &  &-13.70929  &  -13.705    \\
\hline
\hline
$\chi^{xxxx}_{01,01,02,02} $ &   & -19342.219  &   -19401.9   \\
$\chi^{xyxy}_{01,01,02,02} $ &   & -96255.978  &   -96202.6  \\
$\chi^{xxxy}_{01,01,02,02} $ &   & -71907.540  &  -71896.1  \\
$\chi^{xyxz}_{01,01,02,02} $ &   & -113777.75  & -113727.9   \\
\hline

\end{tabular}
\caption{Comparison between analytical and numerically extracted (SVD)
values of the LJ force constants in the diamond structure, when invariance 
relations \ref{trans} and \ref{rot} were not imposed,
and only 8 neighbor shells were included in the fitting. Real interactions were included up to 10 neighbors ($R_{\rm cut}=1.6$). Subscripts in the first column
refer to first or the second atom in the primitive cell ($\tau$ index for $\Phi$ and 
$R\tau$ index for $\Psi$ and $\chi$ ), while superscripts refer to the
cartesian coordinates ($\alpha$ index).} 
\label{ljfc}
\end{table}

The same procedure was followed to extract FCs for a real material case: Silicon (diamond).
In this case, the range of FCs is unknown and probably longer-ranged than we can
exactly handle. The range of harmonic FCs was limited to 8 neighbors, that of the
cubic coefficients was limited to 2, and quartic FCs were limited to nearest-neighbor
interactions. First we used a data set, similar to the previous LJ calculation, where
atoms were displaced by 0.008 and 0.016 $\AA$ along all 3 directions in order to minimize
systematic and round-off errors which occur in first-principles calculations.
LDA-based ultrasoft pseudopotentials were  used within the VASP density functional
simulation package\cite{vasp}. A cutoff energy of 300 eV and a single K-point
(1/4,1/4,1/4) was used in the  $3 \times 3 \times 3$ cubic supercell of 216 atoms.

This was the largest system we could handle. Perhaps more accurate determination
of forces using more K-points and a higher cutoff energy would give better results.

In table \ref{si-compare} harmonic FCs are compared to a recent calculation\cite{qteish} based on 
density-functional perturbation theory\cite{baroni,gonze}.
 
\begin{table}
\begin{tabular}{c|ccc}
FCs & {Nb-Shell} &  Aouissi et al.\cite{qteish} & {SVD} \\
\hline
$\Phi^{xx}_{1,1}$ & 0 (0,0,0) & 13.445  &  13.264      \\
\hline
$\Phi^{xx}_{1,2}$ & 1 (${1 \over 4},{1 \over 4},{1 \over 4}$) & -3.2796  &   -3.2269    \\
$\Phi^{xy}_{1,2}$ & 1 & -2.2875  & -2.2540  \\
\hline
$\Phi^{yy}_{1,1}$ & 2 (${1 \over 2},{1 \over 2}$,0) & -0.1827 &   -0.1873  \\
$\Phi^{yz}_{1,1}$ & 2 &  -0.1768 &  -0.1788     \\
$\Phi^{xy}_{1,1}$ & 2 &  0.1078 &  0.1070     \\
$\Phi^{xx}_{1,1}$ & 2 &  0.4177 &  0.4065   \\
\hline
$\Phi^{xx}_{1,2}$ & 3 ($-{1 \over 4},-{1 \over 4},-{3 \over 4}$) & 0.0321  &  0.0324   \\
$\Phi^{xy}_{1,2}$ & 3 &  -0.0340 &  -0.0354    \\
$\Phi^{xz}_{1,2}$ & 3 &  0.0272 &  0.0283    \\
$\Phi^{zz}_{1,2}$ & 3 &  0.0068 &  0.0073    \\
\hline
$\Phi^{xx}_{1,1}$ & 4 (1,0,0) & -0.019  &   -0.0239  \\
$\Phi^{zz}_{1,1}$ & 4 &  -0.0058  &   -0.0072    \\
\hline
$\Phi^{xx}_{1,2}$ & 5 (${3 \over 4},{3 \over 4},{1 \over 4}$) & -0.0175  &   -0.0197    \\
$\Phi^{xy}_{1,2}$ & 5 & -0.0272  & -0.0265   \\
$\Phi^{xz}_{1,2}$ & 5 &  0.0554  &  0.0573    \\
$\Phi^{zz}_{1,2}$ & 5 & -0.175  &   -0.175   \\
\hline

\end{tabular}
\caption{Comparison between harmonic force constants of Si. 
Again invariance relations \ref{trans} and \ref{rot} were not imposed in the fitting, but
were included by correcting $\Phi_{11}^{\alpha \beta}$. 
Up to 8 neighbor shells were included in the fitting. For brevity, we report the results
on the first 5 shells. Units are in $eV/\AA^2$ }
\label{si-compare}
\end{table}

The phonon spectrum obtained by including harmonic force constants up 
to 8th neighbors is plotted in Fig. \ref{si-phonons} versus experimental
data points.

\bfig[h]
\includegraphics[angle=270,width=0.5\textwidth]{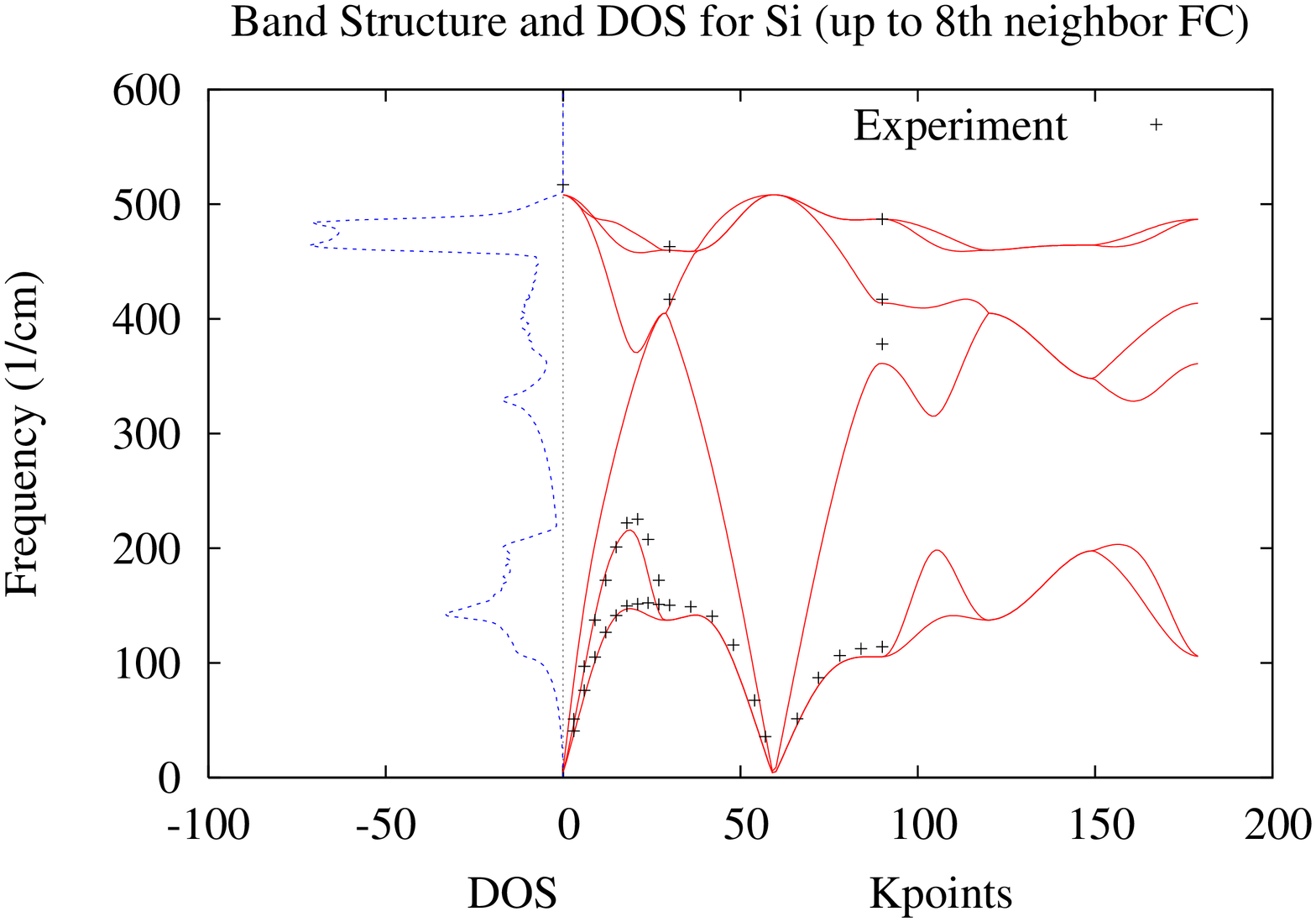}
\caption{Phonon band structure of Si using up to 8$^{th}$ neighbor FC.}
\label{si-phonons}
\efig

\section{Discussions}

The test example of Lennard-Jones potential shows that provided the actual calculation 
of force-displacements is extremely accurate, it is possible to extract harmonic, cubic and
quartic force constants with relatively good precision, by using the above method.
Choosing low-symmetry supercells may be advantageous and avoids heavy calculations 
in big supercells. Our experience on Si has shown that it is possible to extract FCs of Silicon
up to the fifth neighbor by using a small 12 atom ($1 \times 2 \times 3$) supercell. But
longer-range FCs would require larger cells because the fitting would fail in this case.
In fact, large supercells are really needed for the extraction of longer-range harmonic FCs.
Once the latter are known, it is possible to freeze them and extract higher-order FCs from
a smaller size supercell but more accurate calculation.
In this case, the force-displacement data can come from a molecular dynamics run
of small amplitude, as more than one particle is needed to move in order to extract
three and four-body terms.

In case the so-developed potential is going to be used for performing large-scale 
molecular dynamics (MD) simulations, one does not even need to include harmonic (and anharmonic)
FCs beyond a few, say 4 or 5, neighbor shells. In fact most classical force fields have 
shorter range. Although MD runs might be more time-consuming with this method,
it has the advantage of higher accuracy. Lattice distorsions and defects can also
be treated using this potential, as the new force constants of the new atomic
configuration can be obtained from the already developed Taylor expansion:
\beqnar
\Phi_{new}&=&\Phi + u_0 \Psi  + u_0 u_0 \chi /2 \nonumber \\
\Psi_{new}&=&\Psi + u_0 \chi   \nonumber \\
\chi_{new}&=& \chi    
\eeqnar 
in which the static displacement $u_0$ is due to external forces, and
would be obtained by solving $F(u_0)=0$ where the force $F$ is calculated from
Eq. \ref{force}. 

An additional complication arises when one is dealing with polar or
ionic materials. In such cases Born effective charges need to be calculated 
and their long-range effect on the FCs calculated separately using the 
Ewald summation or multipole expansion methods. Care must be taken since 
its short-range part is already included in the extracted FCs. 
So it needs to be properly 
subtracted from the Ewald sum contribution. Needless to say, this 
problem is present in all polar materials, and can not be bypassed.
 

Using the extracted cubic force constants, one can calculate phonon lifetimes and 
frequency shifts. Details of such calculations  will be published elsewhere.

\section{Conclusions}

To summarize, we have developed a method to extract harmonic, cubic and quartic force constants
of any crystal from first-principles force-displacement data. The methodology uses symmetries
of the crystal to reduce the number of independent FCs, and can include
up to any number of neighbor shells in principle. It requires, however very accurate 
first-principles data, in order to produce reliable FCs. This method paves the way
for the development of a new generation of interatomic potentials of ab-initio accuracy.

KE would like to thank the supercomputer center of the IMR, Tohoku University for 
generous supercomputer time, financial support by the University of California 
Energy Institute, and useful discussions with Dr. J. Feldman.

\end{document}